\documentclass[9pt,twocolumn,twoside]{opticajnl}
\journal{opticajournal} 
\setboolean{shortarticle}{true}

\usepackage{wasysym}
\usepackage{bm}

\title{Lithium Faraday Filter: Some Like It Hot}

\author[1]{Maximilian Luka}
\author[1]{Yijun Wang}
\author[1]{Denis Uhland}
\author[1,*]{Ilja Gerhardt}

\affil[1]{light \& matter group, Institute for Solid State Physics, Leibniz University
Hannover, Appelstrasse 2, D-30167 Hannover, Germany}
\affil[*]{ilja.gerhardt@physics.uni-hannover.de}

\begin{abstract}
Magnetically induced rotation of linearly polarized light near an atomic resonance, combined with Doppler-broadened absorption windows, enables narrowband transmission of optical frequencies. An ultra-narrowband lithium vapor Faraday filter at about 671~nm is investigated experimentally and theoretically. The resulting Faraday filter transmittance is demonstrated using a lithium heat pipe oven under longitudinal magnetic fields ranging from 0 to 300~G. Optimization of the lithium Faraday filter performance reveals an optimal operating point at 264~°C and an external magnetic field of 269~G, yielding a peak transmission of approx. 82~\%. The lithium \textit{D}$_1$- and \textit{D}$_2$-transitions are only 10~GHz apart and temperature broadening leads to an overlap of the isotopes \textit{D}-lines. Thus, the applied theoretical model needs to consider both transitions simultaneously. For this purpose, we extended an existing Python library (ElecSus), which now allows for the calculation of the atomic susceptibilities of lithium.
\end{abstract}
\setboolean{displaycopyright}{false} 
\begin{document}
\maketitle

\section{Introduction}
The detection of light underpins a wide range of applications in quantum optics and optical sensing. These experiments rely on near-unitary optical transmission and a well-defined interaction cross-section at a selected frequency to maximize the signal-to-noise ratio. However, achieving such ideal conditions in practice remains challenging, often limiting overall detection efficiency and sensitivity. Atomic vapors can act as a robust and efficient absorbing medium. Close to atomic resonances, the refractive indices change dramatically, affecting the light's transmittance through the atomic vapor. In contrast to solids and liquids, the reduced density in the vapor phase leads to less spectral broadening~\cite{atkins_2008}. In crystals, structural defects serve as an additional source of inhomogeneous broadening~\cite{orth_1993}. Thus, no commercial dichroic filter can compete with bandwidths in the GHz range.

Since the introduction of the Lyot filter in 1933~\cite{lyot1939study}, there has been an increasing interest in the development of new techniques for optical filtering. Yngve Öhman built a narrow-band atomic filter~\cite{ohman_soa_1956}, based on the so-called ``Macaluso-Corbino effect'', evoking rotation of polarized light in the vicinity of an atomic resonance~\cite{macaluso_inc_1899}. Later, the more general term ``Faraday effect'' gained popularity, describing the rotation of linearly polarized light in the presence of a longitudinal magnetic field in any medium~\cite{faraday_trs_1846}. The hereby baptized ``Faraday filter'' transmits only light frequencies in the proximity of an optical atomic transition~\cite{macaluso_inc_1899}. Further research on Faraday filters by Dick and Shay established the term Faraday anomalous dispersion optical filter (FADOF)~\cite{dick_ol_1991}. In contrast, recent studies report that the point of best performance for many Faraday filters does not coincide with anomalous dispersion but instead lies within their spectral wings~\cite{zentile_ol_2015}. These findings indicate that anomalous \textit{dispersion} generally plays a minor role in the overall transmittance of the filters~\cite{gerhardt_ol_2018}. Nevertheless, anomalous \textit{rotation} remains relevant for line-center operation~\cite{kiefer_srep_2014}.

Today, Faraday filters play a significant role in research and sensing. The common media in use are stable alkali metals such as sodium~\cite{chen_ol_1993, zhang_procieee_1999, harrell_josab_2009, kiefer_srep_2014}, potassium~\cite{yin_oc_1992, dressler_josab_1996, zhang_optcom_2001, harrell_josab_2009}, rubidium~\cite{dick_ol_1991, hu_oc_1993, xue_ol_2012, zielinska_ol_2012, uhland_njp_2023}, and cesium~\cite{yin_ol_1991, menders_ol_1991, yin_ieee_1992, wang_ol_2012}, as well as non-alkali metals such as mercury~\cite{fork_ao_1964}, samarium~\cite{barkov_oc_1989}, bismuth~\cite{roberts_jpb_1980}, and calcium~\cite{chan_ieee_1993}. Hot lithium vapor has so far been employed in laser locking and saturated absorption spectroscopy~\cite{eismann_2011, dutta_2023}, and more recently, in magnetometry~\cite{ishikawa_2021} and modulation transfer spectroscopy~\cite{khalutornykh_2025}. The reason for limited applications lies in the low vapor pressure of lithium, which therefore requires high operating temperatures (above 260~°C, leading to $n_{\mathrm{Li}} > \nobreak10^{8}\,\mathrm{atoms}/\mathrm{cm}^{3}$). This vapor pressure is comparable to that of rubidium and potassium at room temperature~\cite{fischer_phd_2023}. Thus, the focus of employing lithium ensembles lies mainly in the ultra-cold regime (e.g., to examine magneto-optical effects~\cite{franke-arnold_jpb_2001}) rather than hot vapor applications.

In this work, we demonstrate a Faraday filter based on hot lithium vapor operating at a center wavelength of 670.96~nm corresponding to the \textit{D}$_2$ transition of $^7$Li. For the analysis, we extend an existing library that calculates the atomic susceptibilities (ElecSus~\cite{ElecSus}). The small separation of the $D$-line transitions requires a simultaneous mathematical description of the electronic susceptibilities, which are only a few GHz apart. Hence, spectra of the $D$-lines for both isotopes overlap, which requires a combination of the initial separate calculations.

\section{Design of the Experiment}

\begin{figure}[ht]
  \centering
  \fbox{\includegraphics[width=\linewidth]{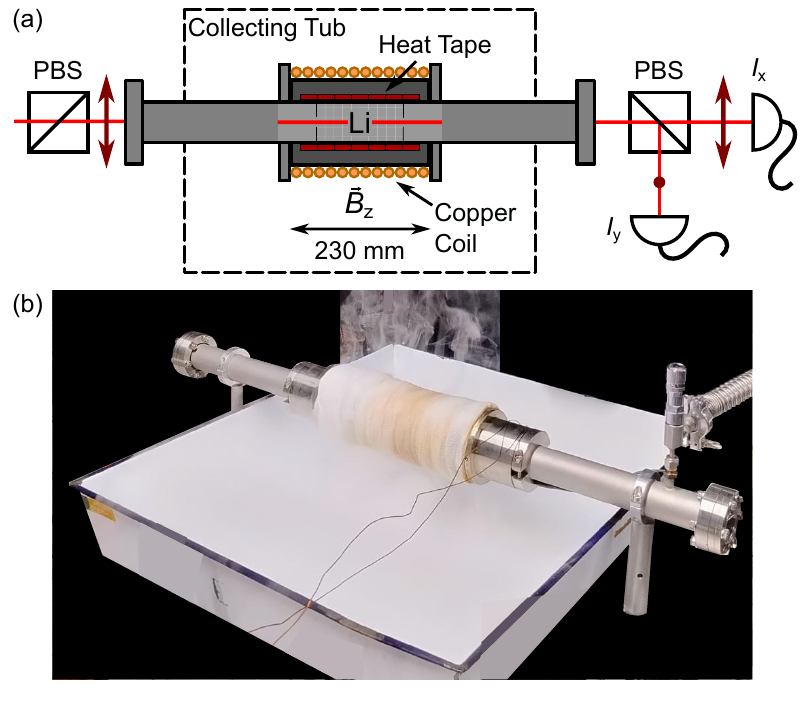}}
  \caption{The setup of the lithium Faraday filter. (a) An evacuated metal pipe with windows is placed between two polarizing beamsplitters (PBS). A cut along the pipe shows the heating tape around the lithium filled cartridge with a nickel mesh. On the outside the copper solenoid is placed to generate an longitudinal magnetic field $\vec{B}_z$ that causes the vapor atoms to rotate the linear polarized light (arrows). Two photodiodes measure the intensity of the rotated light $I_y$ and the unrotated light $I_x$. (b) The experimental setup with the catching container below the steaming hot pipe with a wet bandage. 
  }
  \label{fig:setup}
\end{figure}

A small amount of lithium metal from a commercially available battery is used for our Faraday filter. We place the lithium metal inside a 150~mm self-built steel cartridge with detachable ends with 13~mm holes, centered in an 800~mm long steel pipe with window flanges on each side, as depicted in Fig. \ref{fig:setup}. To prevent lithium oxidation, the system is flushed with argon and evacuated to $10^{-6}$~mbar. A high-temperature heat tape wrapped around the steel tube heats the lithium into the vapor phase. Thermocouples attached close to the heat tape monitored the temperature. A solenoid centered around the heat pipe with an inner diameter of 75~mm, 230~mm length, and 880 windings of insulated copper wire ($\diameter\,1$~mm) generates a longitudinal magnetic field. A small air gap ($\sim10$~mm) is left between the pipe and the solenoid, providing minor thermal insulation against the melting of the copper wire’s insulation. The enormous heat also influences the resistance of the solenoid wire. In addition, the lithium was embedded in a nickel mesh to capture it within the cartridge. The nickel enhances the magnetic field strength. A collection tub beneath the filter is used to collect the cooling water continuously poured over the solenoid. A DC driver (R\&S NGP 800) provides the current for the longitudinal magnetic field along the laser's probe axis. The probing light is derived from a Ti:Sa ring laser (Matisse CS of Sirah, $\Delta\nu\sim$1~MHz), tuned to the $D$-lines of lithium at around 671~nm. Modification to the laser allowed for mode-hop-free scanning of the entire lithium spectrum (60~GHz) at an appropriate speed (1/3~Hz) while simultaneously generating a trigger signal from the piezo scanning cycle. Via an acousto-optic modulator and feedback loop, we stabilize the laser power to 0.2~mW. The beam diameter collimated by an aspheric lens (Geltech, C220TME-B) results in 2.6~mm. This already introduces power broadening, and absorption starts to saturate, such that the medium becomes more transparent. The Rayleigh range is approx. 8.2~m. A simultaneously acquired Fabry-Pérot-spectrum enables the linearization of piezo-induced nonlinearities in the frequency scan. Before and after the heat pipe, we set up polarizing beamsplitters (PBS).

In the presence of an external magnetic field, the atomic levels split due to the Zeeman effect. When linearly polarized light ($\pi$) propagates through the heated vapor, its two circular components ($\sigma^+$ and $\sigma^-$) interact differently with the medium because the Zeeman splitting shifts the atomic resonance frequencies. Thus, each circular component experiences a unique dispersion when interacting with the hot vapor, and the refractive indices become frequency dependent. The two different dispersion relations for the circular $\sigma^+$ and $\sigma^-$ components induce a phase shift that results in a net rotation of the linearly polarized $\pi$ light. Depending on the amount of rotation, the $I_y$ component of the transmitted light reflects from the PBS after the vapor cell. That allows the filter to act as an optical bandpass in the vicinity of atomic transitions. By increasing the magnetic field strength, the Zeeman states can be tuned to create a non-absorbing window at the line center, referred to as line-center operation~\cite{kiefer_srep_2014}. 

Compared to other alkali metals, lithium has the special feature that the $D_1 (2^2S_{1/2} \rightarrow 2^2P_{1/2})$ and $D_2 (2^2S_{1/2} \rightarrow \nobreak2^2P_{3/2})$ transitions are very close to each other ($\sim$10~GHz). The consequences are overlapping spectra for the $D_1$-line of $^7$Li and $D_2$-line of $^6$Li~\cite{fischer_phd_2023, gehm_phd_2003}. Therefore, using the current implementation of the simulation and fitting tool ElecSus was not an option~\cite{ElecSus}, since the computation is based on the choice of either $D_1$ or $D_2$ line~\cite{zentile_cpc_2015}. Handling both optical transitions simultaneously is implemented using calculations based on the atomic Hamiltonian in the completely uncoupled basis. In addition, power broadening and saturation correction of the absorption coefficients are used for all fits~\cite{sherlock_ajp_2009}.  We therefore include the saturation parameter $s=I/I_{sat} = 0.26$ in our analysis, where $I_{sat}$ is the saturation intensity and $I$ the probe intensity.

\section{Results}

To determine the optimal operating conditions for the Faraday filter, we first simulate the spectra for a $\pm$30~GHz detuning centered around the lithium $D$-transitions, extracting possible values for cell temperature, magnetic field, and optical path length. All three parameters are dependent on each other, such that it is difficult to achieve near 100~\% transmission. Based on simulations, we selected a magnetic field of 269~G, a calculated vapor temperature of 264~°C, and a cell length of 150~mm, corresponding to a winged operated filter setting~\cite{zentile_ol_2015}. The vapor temperature sets the density of optically interacting atoms in the optical path.

\begin{figure}[ht]
  \centering
  \fbox{\includegraphics[width=\linewidth]{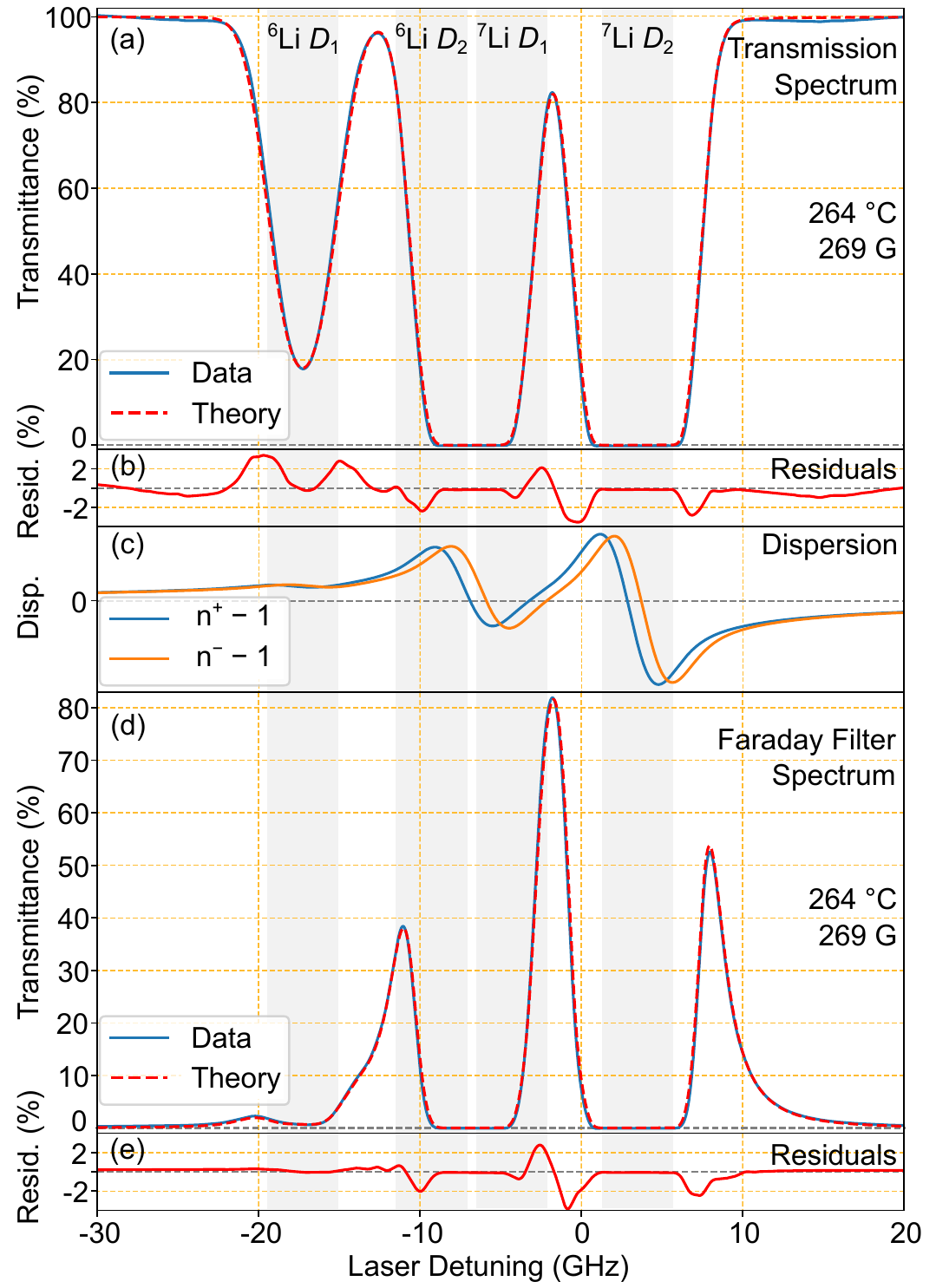}}
  \caption{Transmission spectrum and Faraday filter spectrum of lithium. The light gray background serves only as a visual guide. (vapor temperature = 264~°C, Doppler temperature = 417~°C, length = 150~mm, magnetic field = 269~G) (a) The transmission spectrum is reconstructed from both photodiodes with all $D$-lines of Lithium 6 and 7. Light on the $^6$Li-transitions is not fully absorbed because of the natural abundance of 7.58~\% and the high probe intensity. (b) and (e) The calculated residuals for the transmission spectrum and for the Faraday filter spectrum are lower than ±4~\% at the steep edges. (c) The calculated dispersion for the corresponding parameter set. (d) The Faraday filter spectrum with a maximum transmittance of approx. 82~\%.}
  \label{fig:filter}
\end{figure}

The sum of both recorded components ($I_x$ and $I_y$) reconstructs the Lithium transmission spectrum under the influence of a magnetic field (see Fig. \ref{fig:filter}~(a)). The transmittance is scaled to account for a system without losses from the filter windows. Within our selected scanning range, we cover all lithium $D$-transitions of both isotopes, $^6$Li and $^7$Li. The light corresponding to the $^6$Li transitions is not completely absorbed due to the natural abundance of \mbox{\( {}^{6}\mathrm{Li}/({}^{6}\mathrm{Li}+{}^{7}\mathrm{Li}) = 7.58\,\% \)}~\cite{coplen2002} at the temperature of 264~°C.  That is directly visible for the $^6$Li~$D_1$-transition. The absorption is reduced by $\sim$10~\% due to the high probe intensity, and thus, saturation correction of the absorption is required. Comparing the transmission spectrum to theory shows residuals below $\pm$4~\%, occurring mainly at steep edges (Fig. \ref{fig:filter}~(b)). Based on the transmission spectrum, we calculate the dispersion of the light traversing the lithium vapor (Fig. \ref{fig:filter}~(c)). The lifted degeneracy of the lithium hyperfine states makes the transitions selective to either of the circular components of the linearly polarized probe light ($\sigma^+$ or $\sigma^-$). The difference between the two dispersion relations is proportional to the resulting rotation of the polarized light, leading to the Faraday filter spectrum (Fig. \ref{fig:filter}~(d)). For the given experimental settings, the filter transmits approx. 82~\% close to zero detuning of the $^7$Li~$D_2$ with an equivalent noise bandwidth (ENBW) of 5.32~GHz, and a figure of merit (FOM) of 0.154~GHz$^{-1}$~\cite{zentile_ol_2015}. Despite the lightweight lithium atom and the high Doppler broadening the results are comparable to other Faraday filters based on alkali metals~\cite{gerhardt_ol_2018}.

\begin{figure}[ht]
  \centering
  \fbox{\includegraphics[width=\linewidth]{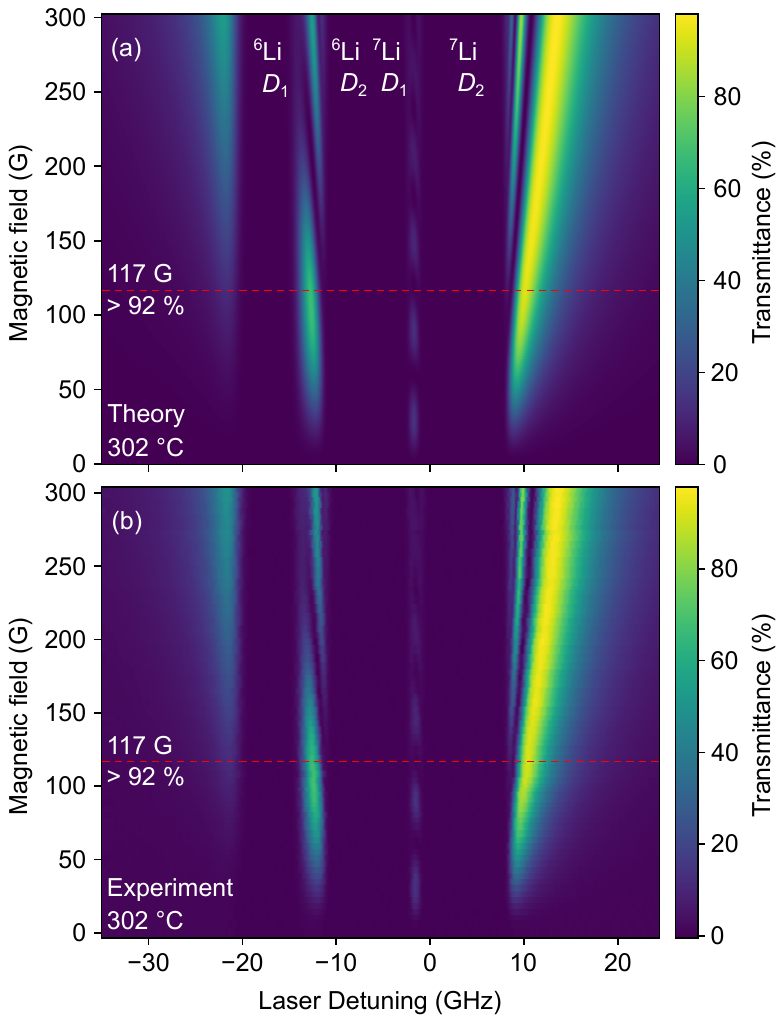}}
  \caption{Density plot of the Faraday filter transmittance for a temperature of 302~°C and optical path length of 150~mm for various magnetic fields. The Doppler temperature corresponds to 505~°C. The red dotted line indicates a spectrum with a peak transmission around 92~\%. (a) The simulated data. (b) The experimental data.}
  \label{fig:spectra}
\end{figure}

In addition, we recorded several Faraday spectra for different magnetic fields from 0 to 300~G, while the temperature and optical path length remained constant. This is consistent with the results extracted from our simulations. In Fig. \ref{fig:spectra}, the resulting density plot of the simulation and the measured Faraday spectra are displayed. With a rising magnetic field, the hyperfine structure splits as predicted by the Breit-Rabi diagram. On the right wing, close to the $^7$Li~$D_2$ transition, multiple rotations arise. Here, the transmittance can exceed 90~\%. For a magnetic field of 117~G (red dotted line), the measured value of the transmittance reaches 92~\%. However, this configuration yields an ENBW$=6.464~\mathrm{GHz}$ and FOM$=0.143~\mathrm{GHz}^{-1}$. The FOM represents the ratio of maximum transmission to the achievable signal-to-noise ratio under white-light illumination, and optimal filter performance requires maximizing this value. Despite the higher transmission, the filter exhibits reduced overall performance in this setting. Rotations of $180$° of linearly polarized light are visible on the wing between $^7$Li~$D_1$ and $^7$Li~$D_2$ around $-1~\mathrm{GHz}$ detuning.

A key experimental challenge in atomic spectroscopy is precisely controlling the desired temperature of the atomic vapor cell. Often, a mismatch between target and measured temperature is observed~\cite{agnew_2024}. The particular case of lithium requires a heat pipe to achieve high enough temperatures. Subsequently, the temperature deviation between the heating tape and the desired atomic vapor temperature is significant, determined to be around 100 K. This also makes it difficult to quickly adjust the temperature, especially to the exact optimum operating point of the lithium Faraday filter. 
The solenoid has poor thermal isolation from the main assembly. That also causes gradual heating of the system and a change in the electrical resistance within the thin copper wire, which affects the magnetic field amplitude. To minimize this effect, a current of 6~A is applied to the solenoid after each measurement sequence to reheat the system. The current is reduced to the target value during data acquisition to maintain steady state conditions. To avoid damage to the solenoid insulation, the experiment was operated at high current only for a limited time. Due to the design of the heat pipe, the internally measured Doppler spectrum shows substantial Doppler broadening, which corresponds to 417~°C in Fig. \ref{fig:filter} and to 505~°C in Fig. \ref{fig:spectra}. This is likely caused by the 13~mm aperture of the inner cartridge, such that longitudinal velocities are increased. An additional broadening may stem from pressure broadening caused by the buffer gas argon~\cite{degraffenreid_2003}, which was not accounted for in this study.

In the theoretical description of the acquired spectra, we observed a strong correlation between the cell length—assumed as the length of the inner region of the lithium-filled cartridge— and the vapor temperature. That arises from the absence of clearly defined cell boundaries and hinders an accurate determination of the effective optical path length, $L$, which is directly affected by the vapor temperature $T$. Their relation is characterized by $\theta_F=\frac{\pi}{\lambda}(\chi^{\prime}_{+}-\chi^{\prime}_{-})L\propto \bm{n(T)L}$ and $L\propto\frac{1}{n(T)}\sim exp(\frac{E}{k_BT})$, which describe the effective rotation $\theta_F$ of the polarized field, where $\chi$ denotes the susceptibilities, $n(T)$ is the atomic vapor density, and $E$ the energy needed to enter the vapor phase~\cite{harrell_josab_2009}. Based on the experimental curves in Fig. \ref{fig:filter} and \ref{fig:spectra}, we chose an optical length of 150~mm and a constant temperature for fitting, despite the absence of sharp cell boundaries and the presence of temperature gradients. These relationships may appear somewhat arbitrary; however, modifying them does not allow the Doppler and vapor temperatures to be matched, and they remain consistent with the physical conditions.

A Faraday filter based on atomic lithium was presented. We utilized a heat pipe to sustain the high temperatures required. Faraday filter spectra with high transmission were measured, as predicted by the theory. It was necessary to implement the theoretical parts. Up to our study, it was only possible to perform calculations with distinguished $D_1$- or $D_2$-lines. Due to the high power of the probe laser, the simulation of the spectra requires accounting for power broadening and saturation correction of the absorption coefficients. The altered program is publicly available~\cite{maximilianlam_repo}. This lithium Faraday filter enables narrowband light filtering at 671~nm. It enhances signal-to-noise performance for high-fidelity optical detection, a capability that is particularly important to improve state readout and operational fidelities in quantum-optics experiments~\cite{blodgett_2023}.

\begin{backmatter}


\bmsection{Disclosures} The authors declare no conflicts of interest.

\bmsection{Data Availability Statement}  Data underlying the results presented in this paper are available in Ref.~\cite{data}.

\end{backmatter}



\end{document}